\documentclass[]{spie}  

\usepackage{amsmath,amsfonts,amssymb}
\usepackage{graphicx}
\usepackage[colorlinks=true, allcolors=blue]{hyperref}

\title{Overcoming temperature limitations in laser cooling using dressed states and diamond vacancies}

\author[a]{Paul Eastham}
\author[a]{Conor Murphy}
\author[a]{Luisa Toledo Tude}
\affil[a]{School of Physics, Trinity College Dublin, Dublin 2, Ireland}

\begin{document} 
\maketitle

\begin{abstract}
The established approach to laser cooling of solids relies on anti-Stokes fluorescence, for example from rare earth impurities in glass. Although successful, there is a minimum temperature to which such a process can cool set by the electronic level spacing in the impurity. We propose an alternative method which does not suffer from this limitation. Our approach relies on the formation of dressed states under strong laser driving, which generates a spectrum in which the gaps can be tuned to optimize the heat absorption. This allows for a cooling cycle which operates at any temperature with a power comparable to the maximum dictated by thermodynamic principles. While this cooling cycle will compete with heating due to non-radiative decay and other mechanisms, it could in principle allow laser cooling to temperatures which are unachievable with anti-Stokes fluorescence.
\end{abstract}


\section{INTRODUCTION}
\label{sec:intro}  

The possibility of using light to cool solids is a model thermodynamic process and has great technological potential. It is well established~\cite{epstein_observation_1995, seletskiy_laser_2010,melgaard_solid-state_2016} that anti-Stokes fluorescence can cool rare-earth doped glasses, with experiments reaching temperatures below 100\,K. However, this approaches the lower limit that is achievable. In this contribution, we describe our recent work~\cite{tude_overcoming_2024} proposing and analyzing a different technique, which we call Dressed-state Anti-Stokes Cooling (DASC), that does not suffer from this limitation. 

\section{DRESSED-STATE ANTI-STOKES COOLING}

The simplest model which allows anti-Stokes cooling is a three-level system, which for definiteness we suppose comprises a single ground state and two excited states. Optical driving of the lowest-energy transition will transfer population from the ground state to the lower excited state. Heat is then absorbed from the host crystal, transferring this population to the upper excited state, and the cycle is closed by radiative decay back to the ground state. This cycle, overall, absorbs one photon from the driving laser and reemits at higher energy with the extraction of heat from the surroundings.  The transition between the excited states involves the absorption of a single phonon, of energy equal to the excited-state splitting $\Delta E$, and occurs with a rate proportional to the number of such phonons. Thus, the cooling power vanishes exponentially below a temperature $kT\sim \Delta E$. 

To avoid this we propose to use strong near-resonant excitation of a dipole-active impurity. The key idea is that phonon transitions then occur between laser-dressed states, whose wavefunctions and energies are modified as occurs in the Autler-Townes/a.c.-Stark effect. This allows the cooling power to be maximized by ensuring that the defect can effectively absorb heat at any temperature. This is expected to occur when there is a transition of energy $\sim kT$, since smaller gaps absorb less energy per cycle, whereas larger ones lead to much slower cycles. 

In a recent paper~\cite{tude_overcoming_2024}, we analyzed this process using a detailed model of the silicon-vacancy (SiV) defect in diamond~\cite{becker_coherence_2017}. This is one of a family of group-IV color centers that are of interest for solid-state quantum optics and quantum technology~\cite{bradac_quantum_2019}. It is a one-hole system with two spin states. The states in each spin sector form a four-level system, with two levels in a ground-state manifold, and two in an excited-state manifold. This structure is directly observed in the photoluminescence, which at low temperature shows four narrow peaks in the spectrum~\cite{hepp_electronic_2014,becker_coherence_2017}. As we showed theoretically, driving on the lowest (or second-lowest) of these lines would allow the implementation of a conventional anti-Stokes cooling process, but this provides only a small cooling power over the temperature ranges of interest, between 1 and 100 K. Analyzing this process in the strong-driving regime, using a Born-Markov approach extended to compute heat flows\cite{murphy_laser_2022}, we found that much larger cooling powers could be achieved due to the level-mixing effects produced by the driving field. The resulting cooling occurs over a range of driving frequencies, making the process potentially robust against inhomogeneous broadening. 

There are, of course, competing heating processes which will make observing cooling using this approach challenging. Non-radiative decay from the excited to the ground state manifolds will produce a strong heating effect which, due to the mismatch of optical and phonon energies, overwhelms the cooling process unless the quantum efficiency is very high. This is also a challenge in anti-Stokes cooling. It is exacerbated here not because of the particular choice of process, but because we seek to cool at low temperatures. As we discuss further in the following, one can argue from thermodynamic principles that there is a maximum cooling power for a single emitter which is on the order of $kT$ per cycle. As our results show, DASC can achieve cooling powers comparable to this bound. However in DASC, and indeed in any optical cooling scheme, a single non-radiative decay for the emitter will produce heat on the order of an optical photon energy, $E_{\mathrm{opt}}$. Non-radiative decay can, then, only occur for fewer than 1 in $E_{\mathrm{opt}}/(kT)$ cycles if cooling is to dominate. At $10\,$K this means that, with an optical energy $E_{\mathrm{opt}}=1$ eV, only 1 in 1000 decays can be non-radiative, and the internal quantum efficiency must be greater than $99.9\%$. 

   \begin{figure}[ht]
   \begin{centering}
       
   \includegraphics{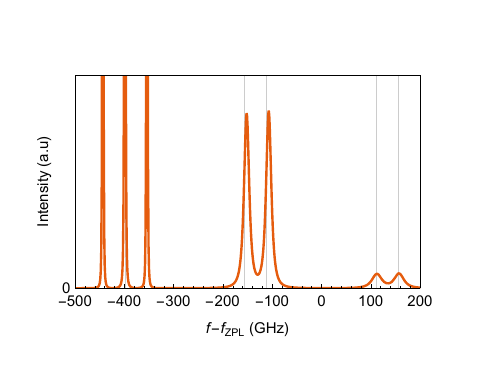}
   
   \end{centering}
   \caption{Computed emission spectrum from an SiV at 20K, reproduced from Ref.\ \citen{tude_overcoming_2024}.  All the three possible polarizations are driven at a single frequency with equal strengths, corresponding to a Rabi splitting of $2\times 10^{-1}\,\mathrm{rad\;ps^{-1}}$. The different polarizations of the output are combined to show the total intensity. Frequencies are measured relative to the zero-phonon line center, $f_{\mathrm{ZPL}}$, and the driving is at $f-f_{\mathrm{ZPL}}=-400$ GHz. The emission frequencies in photoluminescence, corresponding to the transitions in the four-level system, are indicated by vertical grey lines. The spectrum comprises the elastic scattering from the laser, with two sidebands, and four peaks near to the  photoluminescence lines. This spectrum corresponds to inelastic scattering of the driving laser in the strong-driving regime. The intensity on the blue side of the driving laser is greater than that on the red side, implying that energy is being removed from the phonon environment. \label{fig:plspec} }
   \end{figure} 

While very high, we believe that it may be possible for these quantum efficiencies to be reached in diamond vacancies. A low-temperature quantum efficiency of $30\%$ has been inferred~\cite{becker_coherence_2017} from the comparison of the dipole moment, deduced the Rabi oscillations at $5$\,K, and the $\sim 1$ ns decay time of the photoluminescence. However, quantum efficiencies of $67\%$ have been measured at room temperature~\cite{riedrich-moller_deterministic_2014}. There the dominant contribution is expected to be from thermally activated processes which will drop away at lower temperatures, leaving the possibility that quantum efficiencies close to 100\% may be achievable in the absence of confounding material effects such as strain. Measuring these quantum efficiencies directly may be difficult, however, observing cooling or heating would allow them to be inferred. 

The heat extraction in the DASC process could also be demonstrated experimentally, even if it were overwhelmed by a competing heating process, by observing the emission spectrum. Our theoretical methods allow us to compute these spectra and an example, with excitation detuned to achieve cooling, is shown in Figure~\ref{fig:plspec}. It comprises three lines near the driving laser, forming a Mollow triplet structure, and four further lines near those in photoluminescence. This spectrum arises from the scattering of laser photons by the electronic states of the defect. As can be seen, the inelastic scattering is predominantly to higher energies -- the emitted intensity is higher on the blue side of the laser line than on the red one -- demonstrating the extraction of heat from the crystal. 

As well as non-radiative decay there will be competing heating due to absorption of the driving laser in the host. Given our approach involves using fields sufficient to generate laser-dressed states this may be expected to be a particular problem, however our analysis shows that is not the case for the SiV. It imposes a minimum density of SiV needed to achieve net cooling, which we estimate from our model to be $\rho\sim 10^{24}\; \mathrm{m}^{-3}$, taking a conservative value, $\alpha_b=0.1\;\mathrm{cm}^{-1}$, for the background absorption coefficient in diamond. A specific issue for the diamond vacancies is the presence of emission into a local phonon sideband. In order to overcome this heating effect we estimate that one would need a zero-phonon emission fraction of over $0.95$. While larger than the value $0.7$ generally achieved in bulk diamond this appears achievable, with the use of photonic structures to suppress the sideband emission and/or enhance the zero-phonon line.

\section{THERMODYNAMIC LIMITS ON COOLING POWER}

\begin{figure}[ht]
   \begin{centering}
   
   \includegraphics{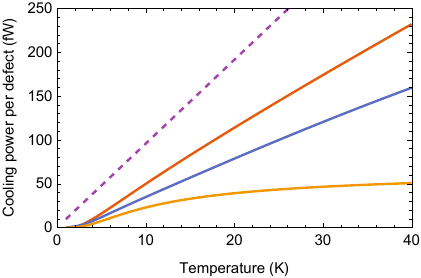}       
   
   \end{centering}
   \caption{Computed cooling power as a function of temperature for the simplified three-level model of the SiV discussed in the main text. The solid curves correspond to driving at a single frequency with strength corresponding to the Rabi splittings $2\times 10^{-0.5}\,\mathrm{rad\,ps^{-1}}$ (red, highest solid curve), $ 2\times 10^{-1}\,\mathrm{rad\,ps^{-1}}$ (blue) and $2\times 10^{-1.5}\,\mathrm{rad\,ps^{-1}}$ (yellow, lowest curve). For each driving and temperature we take the driving frequency which maximizes the cooling power. The dashed purple curve is the thermodynamic upper bound for the cooling power of a two-level system with cycle rate equal to our assumed radiative lifetime, $\gamma=1$\,ns.  \label{fig:3ls} }
   \end{figure}

While in experiments cooling powers are typically limited by competing heating effects, it is interesting to note that thermodynamics leads us to expect a maximum cooling power even in their absence. To see this, we consider an idealized optical cooling process, involving a few-level system which operates cyclically with a rate $\gamma$ to extract heat from a reservoir at temperature $T$ and transfer it, by radiative decay, to an electromagnetic environment. We take the start of the cycle to be the situation immediately after the decay, where the von Neumann entropy $-\mathrm{Tr} \rho \log \rho$ of the few-level system is zero. This holds within the well-justified approximation that we need consider only decay into the electromagnetic environment, and not absorption from it, meaning it is effectively at zero temperature. The next part of the cycle will involve increasing the entropy of the few-level system by allowing it to interact with the reservoir to be cooled. This can take it no further than the maximally-mixed state, with entropy $\log N$. This is then the largest possible entropy increase of the working medium in the absorption stroke, and the largest possible entropy decrease of the reservoir, which would be reached only if the heat transfer is reversible. The heat absorbed from the reservoir is then $kT\log N$, and combining this with the cycle rate implies a limit on the cooling power $kT\gamma \log N$. 

An interesting feature of our approach is that it achieves cooling powers comparable to this thermodynamic bound. Figure\ \ref{fig:3ls} shows results for the maximum cooling power of a three-level system, driven by a single-frequency field of various strengths, corresponding to the different curves. The calculation is performed with a simplified three-level version of the four-level model we considered for the Si-V. This simplified model is obtained by replacing the two ground states with a single state, and produces qualitatively similar results to the four-level case. For each temperature and driving strength we find the driving frequency that maximizes the cooling power. The cooling power has the same linear temperature dependence as the thermodynamic bound obtained for a cyclic process, shown for a two-level system as the dashed line, and indeed is of a comparable magnitude. 


\acknowledgments 

We acknowledge funding from the Irish Research Council under award GOIPG/2017/1091, and Science Foundation Ireland (21/FFP-P/10142). In order to meet institutional and research funder open access requirements, any accepted manuscript arising shall be open access under a Creative Commons Attribution (CC BY) reuse licence with zero embargo.

\bibliography{references} 
\bibliographystyle{spiebib} 

\end{document}